# Using Elo Rating as a Metric for Comparative Judgement in Educational Assessment

ANDY GRAY, SWANSEA UNIVERSITY, WALES

Swansea University, Swansea, Wales, 445348@swansea.ac.uk

ALMA A. A. RAHAT, SWANSEA UNIVERSITY, WALES

Swansea University, Swansea, Wales, alma.a.a.rahat@swansea.ac.uk

TOM CRICK, SWANSEA UNIVERSITY, WALES

Swansea University, Swansea, Wales, tom.crick@swansea.ac.uk

STEPHEN LINDSAY, GLASGOW UNIVERSITY

Glasgow University, Glasgow, Scotland, Stephen.lindsay@glasgow.ac.uk

DARREN WALLACE, CDSM, WALES

CDSM, Swansea, Wales, Darren.wallace@cdsm.co.uk

*M*arking and feedback are essential features of teaching and learning, across the overwhelming majority of educational settings and contexts. However, it can take a great deal of time and effort for teachers to mark assessments, and to provide useful feedback to the students. Furthermore, it also creates a significant cognitive load on the assessors, especially in ensuring fairness and equity. Therefore, an alternative approach to marking called comparative judgement (CJ) has been proposed in the educational space. Inspired by the law of comparative judgment (LCJ). The key idea here is that the better submission between a pair will be identified by a suitably qualified or experienced assessor. This pairwise comparison for as many pairs as possible can then be used to rank all submissions. Studies suggest that CJ is highly reliable and accurate while making it quick for the teachers. Alternative studies have questioned this claim suggesting that the process can increase bias in the results as the same submission is shown many times to an assessor for increasing reliability. Additionally, studies have also found that CJ can result in the overall marking process taking longer than a more traditional method of marking as information about many pairs must be collected.

There is a clear necessity to investigate the efficacy of alternative rating and ranking systems that do not require extensive data on every pair of submissions, to reduce the temporal and cognitive burden on assessors, and bias from observing the same submission repeatedly. In this paper, we investigate Elo, which has been extensively used in rating players in zero-sum games such as chess – for devising a ranking between submissions in a comparative judgement context. We experimented on a large-scale Twitter dataset on the topic of a recent major UK political event ("Brexit", the UK's political exit from the European Union) to ask users which tweet they found funnier between a pair selected from ten tweets. Our analysis of the data reveals that the Elo rating is statistically significantly similar to the CJ ranking with a Kendall's tau score of 0.96 and a p-value of $1.5 \times 10^{-5}$. We finish with an informed discussion regarding the potential wider application of this approach to a range of educational contexts.

**ACM Reference Format:**
Andy Gray, Alma A. A. Rahat, Tom Crick, Stephen Lindsay, and Darren Wallace. 2022. The Title of the Paper: ACM Conference Proceedings Manuscript Submission Template: This is the subtitle of the paper, this document both explains and embodies the submission format for authors using Word.

## 1 INTRODUCTION

The impact of the COVID-19 global pandemic on all educational settings and contexts is profound and still not over; we will likely experience a "new (ab)normal" over the coming period [28], especially with the longer term impact of digital education, pedagogy and practice [27]. We have thus seen the significant impact on learning, teaching and assessment, across all subjects and disciplines [8, 21]; even taking into consideration some of the benefits that may be realised from rapid changes to the delivery of education [7]. However, with one in three school-level teachers [29] looking to leave the profession within the next five years in the UK, and while we appreciate the ongoing impact of COVID-19 has not helped, it is stated that teachers' workload is one of the significant drivers. A National Education Union poll in the UK concluded that 70% of teachers are concerned about the increased workload, and 51% of teachers who intended to leave teaching said so because of the workload [29].

The process of marking within teaching and learning, while crucial, is a time-consuming task for any teacher. A teacher marking a class assessment for 30 students can take multiple hours, and given other work commitments, it would likely require a few days to complete. This might not seem extreme on the surface. To put it into context, let us consider a teacher's full workload [26]; a teacher often only gets 10% planning, preparation and assessment (PPA) time in the UK [17]. Now, if a teacher has 11 classes with 30 students, then 330 pieces of work will need to be marked. In some schools, the marking policies require teachers to assess the students' work every two weeks. That means that for the *39-week academic year*, teachers are expected to mark *330 pieces* of student work *19.5 times*. Therefore, a single teacher may need to mark approximately *6,500 individual pieces of student work* over a single academic year. This number is not taking any mock or additional assessments into account so that these additional marking would be more on top again for the teacher to mark. This is certainly overwhelming, and it is unsurprising that teachers are considering leaving their job. Additionally, with teachers marking individual students, potentially over several days, the process can lead to potential teacher inconsistencies with the marking, and potential unintentional bias becomes present within the results. The unintentional bias could be due to teachers potentially taking into account the students' performance over the whole year rather than the individual assessment or test. Clearly, we need to modernise how we assess and mark individuals.

A key alternative method to traditional marking is a comparative judgment (CJ) technique. The key idea is to present a pair of submissions to an assessor, and then the assessor decides which one is better. The assessor would have to perform such comparisons for as many pairs as possible. This information can then be used to derive a ranking of the submissions. However, while CJ does solve two critical issues regarding teachers' cognitive loads and teacher bias, it does carry its flaws. Including increasing the length of time to complete the overall marking task. Additionally, the marking process also biases the results due to the possibility of repeatedly seeing the same submission, but in differing contexts.

It should be noted that an adaptive version of CJ is currently used in the industry [15, 20]. Adaptive comparative judgement was first proposed by Pollitt in 2011, with TAG Assessments, a company providing evidence-based assessment solutions, developing the initial system [16]. How ACJ gets conducted is that it takes several judges, which in our context would be the teachers marking the students' work, and then provides them with random pairings of work to compare on its initial run-through. Once completion of the initial run, the judges are then presented with more work to be compared. However, this time around, similarly scored pieces of work get compared more often to determine the rank order of the work. Using statistical quality controls to help determine the reliability of its rankings is why many judging rounds are required to allow the statistical quality controls to have enough information to be confident in the model's forecasts. Studies have shown that ACJ takes longer [2, 5] to conduct that standard marking and creates its own biases [5]. A number of companies now offer ACJ tools, for example, RM Compare, a consortium of universities called D-PAC and No More Marking [31]. In this paper, our aim was to perform CJ in a reasonable amount of time. As such we exclude ACJ from further exploration.

Through addressing these issues, the main contributions of this paper are as follows:
- We propose an alternative ranking method to score the results to a CJ assessment, namely Elo, a score frequently used in chess competitions to rank players and estimate the potential winner of a match.
- We compare the proposed Elo system and a traditional CJ on a real dataset, and show that it is a viable alternative.

The rest of the paper is structured as follows; in Section 2, we present the background of the study. Section 3 presents the related work that has been carried out, with Section 4 outlining the method used to create the novel approach to ranking marked student work. We give our results and discussions in Section 5, with a general conclusion expressed in Section 6.

## 2 BACKGROUND

### 2.1 Teaching and Assessment

While there can be multiple reasons why educators assess students, assessments aim to serve a purpose to both the teacher and the student in the process. These include: giving feedback to teachers and learners; providing motivation and encouragement; boosting the pupils' self-esteem; a basis for communication; a method to evaluate a lesson, training method, scheme of work, or curriculum; to entertain the students [30]. Additionally, the assessment process creates opportunities to rank students, ultimately allowing schools to select and filter students, allocate students a particular pathway or educational direction, or discriminate between students for a given set reason [30].

There are four main categories of assessment in the UK. These are diagnostic, formative, summative, and national assessments [9, 30]. However, it is essential to note that national assessments do not get used within everyday aspects of teaching and learning. The term national assessment is used to represent the critical examinations like SATS, GCSE, and A-level examinations taken at the end of the qualifications.

Teaching is becoming increasingly target-oriented and evidence-based in the UK. For example, the Office for Standards in Education (Ofsted) in England carry out inspections and conduct teacher performance reviews. Therefore, there is an emphasis on rigour, and consequently, teachers are encouraged to provide written summative feedback. While verbal feedback is proven to be just as effective to students learning and feedback, it is difficult to record and evidence its efficacy to governing bodies. Therefore, a teacher will mark students' work based on a rubric of key criteria that match certain levels to produce a grade and then provide personalised feedback to a student, explaining what they have done well and what they need to improve on. However, this approach to marking is very time consuming and generates a substantial cognitive load for the teachers. Additionally, with this marking getting done over several marking sessions, this might result in the teacher not being consistent with their marking and feedback. Furthermore, the teacher might inadvertently introduce personal bias while marking the students' work due to the teacher considering how the student has performed all year round, rather than face value in the assessment.

**2.2 Comparative Judgement**

Let us take a teacher who is expected to mark a controlled assessment of several students. The goal here is to assess where a student is currently regarding their expected progress. However, the teacher's approach is to take the current marking rubric provided by the examination board and then mark each student independently, in absolutes, to determine what this particular student had done based on the marking criteria. Each time, this marking process is done for each student, only focusing on one student. The expectation of the school and examination boards for the teacher is to mark each piece of work at face value, ensuring that how the student had done in that controlled assessment was a true reflection and that no previous facts came into play.

An example of this might be that student A is usually a hard-working student within lessons and always does their best when completing tasks. At the same time, student B produces good work but can be inattentive and frequently go off-topic during class lessons. While the marking should be done anonymously, it is challenging for a teacher not to recognise which student's work they are observing. Ultimately, this can lead to bias within the marking and the teacher, especially they are uncertain on what to provide the student. So, the teacher will usually take factors into account, like how the student has performed throughout the year. Therefore, unintentional bias gets added to the results for the students and results in student B receiving better marks than what they should be entitled having.

Comparative judgement (CJ) can be used to alleviate the bias. It is simple in its design but extremely powerful. Louis Leon Thurstone created CJ in 1927, and it was known as the "law of comparative judgement" (LCJ) [23, 24]. Thurstone, a psychologist, created this paradigm when he discovered that human minds are much better at making pairwise comparisons of observed items. A good example of this is: it is easy to compare two objects by weight and then rank a number of objects, rather than ranking by making the judgements in absolutes. The LCJ represents how we perceive things or situations rather than measurements of actual properties [1]. The initial focus of the LCJ was within the psychometrics and psychophysics academic space [11, 12]. Initial examples of the LCJ are:

- Comparing the observed intensity of the weights of objects.
- Allowing the ability to compare the extremity of an attitude expressed within statements, such as statements about capital punishment.
- Asking people what object is more prominent in size.

LCJ is based on comparing a pair of Normally distributed variables. Consider two essays submitted by students $x$ and y. Now, the assessor has a preference over each essay that can be represented as random variables $X$ and $Y$ respectively. Consider that both these variables are Normally distributed, i.e. $X \sim \mathcal{N}(\mu_X, \sigma_X)$ and $Y \sim \mathcal{N}(\mu_Y, \sigma_Y)$. Here, $\mu_X$ and $\mu_Y$ are average preferences, and $\sigma_X$ and $\sigma_Y$ are the associated standard deviations. Since there is a dependency between these preference densities due to the same judge, then the difference between these densities is Normally distributed [4, 25]:

$$X - Y \sim \mathcal{N}(\mu_{XY}, \sigma_{XY}), (1)$$

where, $\mu_{XY} = \mu_X - \mu_Y$ and $\sigma_{XY} = \sqrt{\sigma_X^2 + \sigma_Y^2 - 2\rho\sigma_X\sigma_Y}$ with $\rho$ representing the correlation between the random variables.

This allows us to probabilistically query whether A's work is preferred over B's work. If the preference densities are known, this probability can be computed analytically using the following formula:

$$P(X > Y) = P(X - Y > 0) = \int_0^\infty \phi\left(\frac{t - \mu_{XY}}{\sigma_{XY}}\right) dt = 1 - \Phi\left(\frac{\mu_Y - \mu_X}{\sigma_{XY}}\right) = \Phi\left(\frac{\mu_X - \mu_Y}{\sigma_{XY}}\right), (2)$$

where, $\phi(\cdot)$ and $\Phi(\cdot)$ are the standard Normal probability density and cumulative distribution functions respectively.

With this, we can then rank each individual in a cohort in the order of probabilistic preferences. However, this is a complex problem as there are issues like the estimation of various parameters, adjustments for multiple comparisons [13], etc. Hence, many simplifications and assumptions have been offered in the literature [2, 4, 16, 19, 24, 25].

In this paper, we focus on using an estimation technique proposed by Bradley and Terry (BT) in their seminal paper on the topic [2, 3, 16, 19]. The technique is an iterative minorisation-maximisation (MM) method [14] for estimating the maximum likelihood of the expected preference $\mu_i$ for the $i$th student's essay given the observed data. We can then simply use the expected preferences to sort the essays and generate a rank where a higher value represents a better ranked essay. For achieving this, a series of simplifications of (2) (due to Thurstone [24]) and a logistic transformation is considered, which, for comparing two preference densities $X$ and $Y$, compresses to:

$$\text{logit}(P(X > Y)) = \mu_X - \mu_Y. \quad (3)$$

In this model, we are computing the probability that $x$ wins over $y$ in *log odds unit or logit* where there are only two options: either $x$ wins or $y$ wins [4]. It should be noted that there are other versions that are designed to address three possibilities win, draw or lose; interested readers should refer to [24] for more details.

Consider a vector of average preferences for n essays $\boldsymbol{\mu} = (\mu_1, \ldots, \mu_n)^\top$ with the condition $\sum_i \mu_i = 1$. Let, $\omega_{i,j}$ represents the number of times the $i$th student's essay was preferred over the $j$th student's essay. By definition, $\omega_{i,i} = 0$. The likelihood of the average preference vector $\boldsymbol{\mu}$ under the assumption of the BT model is then given by:

$$L(\boldsymbol{\mu}) = \sum_{i=1}^n \sum_{j=1}^n \left(\omega_{i,j} \ln \mu_i - \omega_{i,j} \ln(\mu_i + \mu_j)\right). (4)$$

The MM algorithm then iteratively updates each $\mu_i$ such that (4) is maximised. The iterative update formula for $k$th iteration is [12]:

$$\mu_i^{k+1} = \Omega_i \sum_{j|j \neq i} \frac{\omega_{i,j} + \omega_{j,i}}{\mu_i^k + \mu_j^k}, (5)$$

Where, $\Omega_i = \sum_j \omega_{i,j}$ is the number of times the $i$th essay has won. At each iteration, we are further required to normalise the $\mu_i$s to ensure that the average preference vector sums to 1:

$$\mu_i^{k+1} \leftarrow \frac{\mu_i^{k+1}}{\sum_j \mu_j^{k+1}}. (6)$$

Overall, under certain assumptions, the iterative process is guaranteed to converge to the optimal $\boldsymbol{\mu}$ [14]. At the final stage, for ease of presentation, we multiply $\mu_i$ by 100 to scale it between 0 and 100.

## 3 ELO RATING SYSTEM

The Elo rating system was invented by Arpad Elo [10]. It was originally designed for generating ranking between chess players in a tournament. The Elo system looks at the difference in the ratings of two players, and then predicts the outcome of a match between them. A player's Elo rating is depicted as a number and will change over time depending on the outcomes of a series of matches (not necessarily from a single tournament), with the winners gaining points over the losers. However, how many points get awarded depends on the difference in ranking between the players. Only a few rating points get taken from the lower-ranked player if the higher-ranked player wins. However, if an 'upset win' occurs, when the considerably lower rank player beats the higher rank player, a much greater number of points will be gained to the winner and deducted from the loser. Ultimately, even when 'upset wins' happen, the ranking of the players will reflect the valid scores over time [22].

The premise for developing Elo score is the same as LCJ, i.e. a player's performance is deemed as a random variable with Normal density that has an expected rating $\mu_i$ for the $i$th player. However, it has a different model for computing the probability in (2) [10, 18]:

$$P(X > Y) = \frac{1}{1+ \exp^{-(\mu_X-\mu_Y)/400}}. (7)$$

Let, the outcome of a match between two players $i$ and $j$ $\mathcal{O}_{i,j} \in \{0,1\}$, where a win for $i$ is represented with 1. With this, we can update $i$th individual's expected performance, or rating as follows:

$$\mu_i \leftarrow \mu_i + K\left(\mathcal{O}_{i,j} - P(i > j)\right), (8)$$

where, K is a coefficient used to control the amount of reward or punishment that can be earned from a single comparison. In chess, this factor is varied based on the experience or rating of a player: for instance, if a player is highly ranked, and they may have a high K factor, so they gain substantial points when they win, and *vice versa*. In the context of comparing essays, this provides us with the opportunity to insert the prior performance of a student. However, we do not consider this in this paper, and use a constant $K = 32$ for all submissions. Like CJ, we use sorting on $\mu_i$s to generate a ranked list, as such a higher numerical value represents a preferred submission.

There is a study comparing Elo and the BT model for rating snooker players [6]. While the study found that both worked well overall, it found that the Elo ranking was better at rating more recent events. Therefore, this would be the same case as students' work getting marked for the first time, instead of over a long period. Additionally, the study suggested that both models were not good when new players entered the league due to their lack of prior performance knowledge [6]. However, this would prove helpful to us as all work will get compared by the participants with no prior knowledge. Therefore, all are seen at the same level with the same opportunity. So, this critical factor in previous studies will not be an issue within ours.

It should be noted that studies propose using Elo within adaptive educational systems (AES) [18], but its execution and purpose are different from what CJ aims to achieve. Within AES, Elo is used as a method to rank the students, to gain their level of understanding of a topic to then decide on what the next question the student should receive, that will be at the appropriate level of their knowledge and understanding of the given topic. Aiming to challenge them but not be beyond their abilities. However, depending on the predetermined perceived difficulty of the question, a higher or lower $K$ value is used to determine the outcome of the student's performance. For example, if a question is perceived to be difficult, then a low K value is used, while if a question is perceived to be easy for the student to answer, then this will have a larger K value. Therefore, impacting on the student's overall score is more significant if the student got the perceived more straightforward question wrong on their score than if they got the tricky question wrong.

While AES is not directly linked to how we propose using Elo scoring to rank the students' work for CJ, it provides insights on applying the K factor to the Elo ranking as the tweets will be of an unknown entity to the markers. Therefore, we have no prior assumptions of perceived skills between the Tweets, so a K value of 32 will get used as the tweets need to be taken at face value of how they score relating to the outcomes of the markers' judgements.

## 4 EXPERIMENTAL SETTINGS

We developed a web interface (Figure 1) that presents two tweets to the participant. The participant can then select the one that they found to be the funnier. We also allowed them to provide us with textual feedback on why they may have thought one was funnier than the other. There was no time limit within which they had to complete the survey.

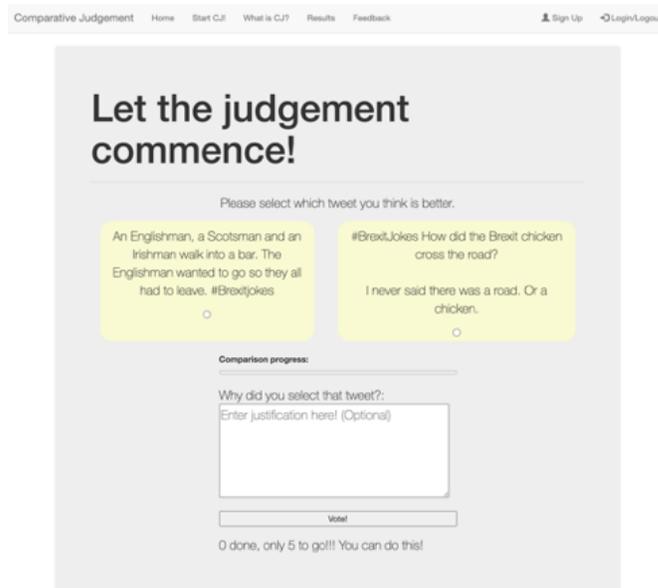

Figure 1: The user interface each user was presented with when taking part in the CJ experiment.

We selected 10 random tweets on the topic of Brexit that were supposed to be funny (see Table 1). The participants were presented with a random combination of these tweets, ensuring that the users only see that tweet once. Therefore, ensuring that the tweet's impact is not lost by the user seeing it multiple times when getting compared. This is particularly important here as a joke may lose its efficacy on a person as we show it to someone repeatedly. As a result, the user only sees 5 different tweet combinations selected from 49 possibilities. We aimed to ensure that we would collect as much data as we can on every possible pair. Although, Elo rating or CJ does not depend on this, we wanted to ensure that we construct a balanced dataset. Multiple users were asked to make the comparisons. With the equations (7) and (8), we computed the Elo score after each comparison that gets made. We used the procedure in equations (5) and (6) to compute the CJ scores.

## 5 RESULTS AND DISCUSSION

In this section, we compare the results against the CJ and Elo ranking. A total of 40 different users participated in the comparison judgement. Through looking at Figure 2 we can see that all combinations were displayed to the participants in this experiment. We can see that the pair of tweets 1 and 5 appeared the most (8 times), while the combination appearing the lowest was the pair of tweets 6 and 7 (only once).

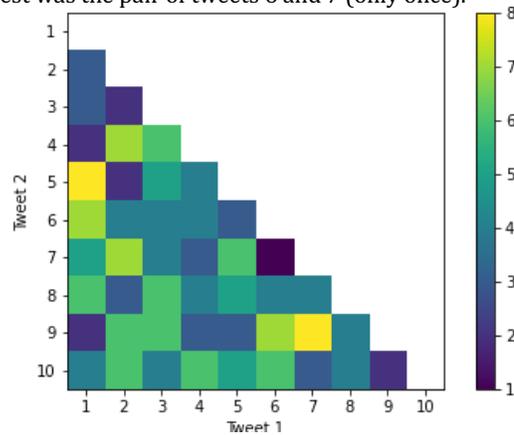

Figure 2: The web applications generated results compared against each other.

As we only wanted a tweet to be shown once to a user and the combinations to be random, our algorithm would generate all the pairings then randomise the order. Once a tweet appeared within a combination, it removed it from any other combination pairings. Therefore, the results show that the method enabled all comparisons to be presented to users at least once, indicating that 40 participants were enough for the data size we used.

When we look at winners and losers of the comparisons (see Figure 3), we can see that the tweet that won the most between a specific combination was tweet 4 and 2, with tweet 4 winning 6 times and tweet 2 winning only once. Additionally, when we look at the combination that appeared the most, 1 and 5, one came out on top 5 times, compared to 5 wins between the two once.

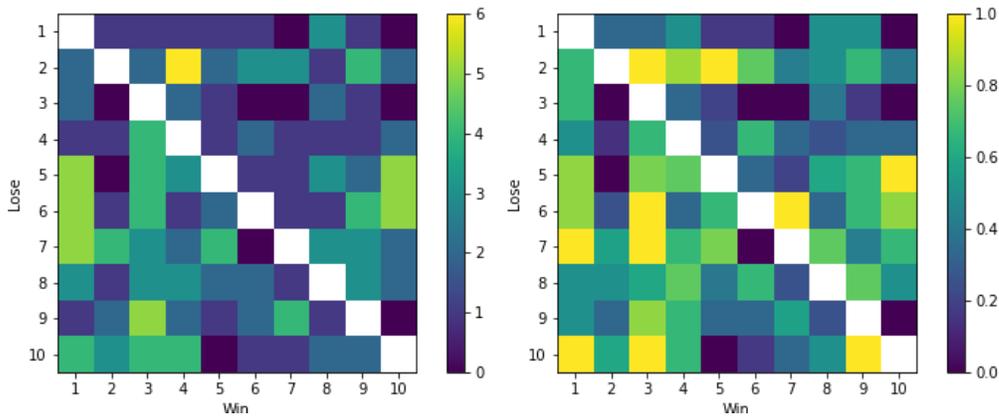

Figure 3: A heat map of the number of times a tweet won or lost. Left - by total values. Right - By win percentages.

When we look at the winner heat map (see Figure 3), we can see that 2, 5, 6, 7 and 10 had moments where they did not win a head-to-head with another tweet. 2, 6, 7 and 10 did not win against at least two different tweets, while the others were only against one tweet they failed to win. We can see that certain tweets never won against another tweet. For example, Tweet 10 never beat Tweet 9, which is also reflected in the ranking of the tweets, as Tweet 9 is ranked higher than Tweet 10 in both the Elo and CJ ranking table. The same can be said about Tweet 6 and 3, with Tweet 6 never beating Tweet 3, resulting in Tweet 6 coming 9th, and Tweet 3 coming 1st in the rankings.

When we look at the two scores plotted against each other, Elo and CJ (see Figure 4), it shows that these values clearly indicate a strong linear relationship. The results returned as 0.96 when a Pearson's correlation test was conducted on these scores. Therefore, the two values are heavily linked, so when a tweet has a good Elo score, it also has a good CJ score. In addition, the Kendall's tau measure on the data reveals that the score is 0.96 with a p-value of $1.5 \times 10^{-5}$. This clearly shows that the results are statistically significant, and therefore, the Elo score is a nearly perfect alternative to the CJ scoring system.

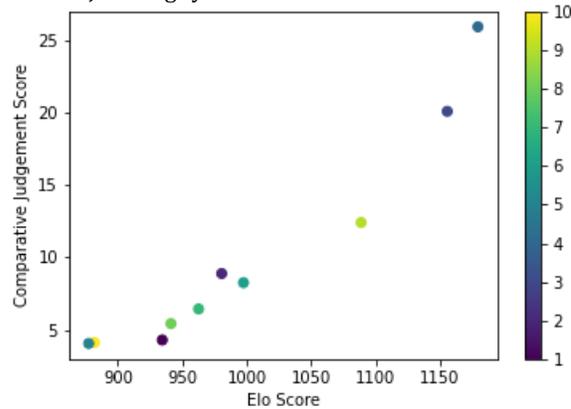

Figure 4: A comparison between Elo and standard comparative judgement scores. The colours represent the tweet IDs. Clearly, there is a strong linear relationship between the scores.

Using the Elo system also provides the process with a lot more ease to compute. It allows the ranking to get done to a high degree of accuracy without the need for iterative approximation. Additionally, the Elo system will work effectively without presenting every combination against each other, which would be useful if the sample size increased. As a result, this would be a sound scoring system to implement at a national scaled-up scale.

While looking at Table 1, we can see that the Elo and CJ ranking generated very similar results. However, as we can see, the tweets coming in 4th and 5th vary slightly between the two ranking methods. These CJ results raise questions about whether further work is required to rank them more accurately, but the CJ ranking is very close. With only 0.63 between the two scores. However, we need to ensure that the process does not end up having someone do multiple rounds and then expand the time required to complete the CJ, taking away any actual benefits. Nevertheless, it does bring to light how effective the Elo ranking system is and can handle these situations and shows it can overcome the potential ambiguity nature of the CJ and have more robust statistical measures in place to generate an overall ranking.

Table 1: The table shows the Elo and CJ scores and ranking position for each tweet, which is identified by its tweet ID and content. At the time of writing, the tweets were publicly available.

| Tweet ID | Tweet | Elo Rank Order | Elo Score | CJ Rank Order | CJ Score |
|---|---|---|---|---|---|
| 3 | *Q: With Britain leaving the EU how much space was created? A: Exactly 1GB* | 1 | 1179.38 | 1 | 25.89 |
| 1 | *An Englishman, a Scotsman and an Irishman walk into a bar. The Englishman wanted to go so they all had to leave. #Brexitjokes* | 2 | 1155.59 | 2 | 20.07 |
| 4 | *VOTERS: we want to give a boat a ridiculous name UK: no VOTERS: we want to break up the EU and trash the world economy UK: fine* | 3 | 1088.82 | 3 | 12.41 |
| 9 | *Hello, I am from Britain, you know, the one that got tricked by a bus* | 4 | 997.55 | 5 | 8.27 |
| 8 | *Say goodbye to croissants, people. Delicious croissants. We're stuck with crumpets FOREVER.* | 5 | 980.64 | 4 | 8.9 |
| 10 | *How many Brexiteers does it take to change a light bulb? None, they are all walked out because they didn't like the way the electrician did it.* | 6 | 962.74 | 6 | 6.46 |
| 5 | *#BrexitJokes How did the Brexit chicken cross the road? "I never said there was a road. Or a chicken".* | 7 | 941.31 | 7 | 5.45 |
| 2 | *Why do we need any colour passport? We should just be able to shout, "British! Less of your nonsense!" and stroll straight through.* | 8 | 934.56 | 8 | 4.32 |
| 6 | *After #brexit, when rapper 50 cent performs in GBR he'll appear as 10.00 pounds. #brexitjokes* | 9 | 881.94 | 9 | 4.15 |
| 7 | *I long for the simpler days when #Brexit was just a term for leaving brunch early.* | 10 | 877.47 | 10 | 4.08 |

## 6 CONCLUSION

The process of CJ aids in reducing cognitive load, as we are generally suited to comparing one thing to another and indicating which one is better [19, 23, 24]. The literature around CJ firmly claims that ACJ is a better alternative to more traditional marking methods, for example, using a rubric. However, CJ does have several flaws. One of the flaws is that the whole process can take longer than traditional marking in the first place. Additionally, the adaptive nature of ACJ can generate bias within its results by getting the markers to mark more often, especially when the results get closely ranked to each other. It gets claimed that a random pairing is better than the adaptive approach.

While CJ generates results to create a ranking of the students' work, CJ is not the only ranking method available; indeed, multiple ranking systems are used within competitive chess and e-sports. In this study we compare two tweets and declare what tweet they preferred. The results were then used to calculate a CJ score and an Elo score, allowing us to compare the final results of the two ranking systems.

The results from the experiment presented that the final Elo ranking and the CJ score a strongly correlated, with a score of 0.96. The web app allowed the users to complete the comparisons quickly and only do one round of judgements. Therefore, reducing cognitive load and reducing the time required for marking. However, the scores only became truly useful after several users had completed the comparison. Still, the more users took part, the final results improved, with the results showing that the Elo system is a suitable method for ranking the results.

Future work on this study will be to apply the approach to real-world student work, researching the outcomes of an Elo rank compared to the grades provided by the teacher. Additional work also needs to be carried out to see if applying Elo ranks to individual learning objectives can generate an overall rank score and provide breakdowns for the students on the areas they did well in and the areas they can improve on. The approach ultimately enables the ability to provide individual feedback to the students based on the marking scores generated. Given the significant impact of the COVID-19 pandemic of traditional assessment, there is significant potential for better understanding and evidencing student performance and achievement across a wide range of educational settings and contexts.

## ACKNOWLEDGMENTS

Andy Gray is funded by the EPSRC Centre for Doctoral Training in Enhancing Human Interactions and Collaborations with Data and Intelligence Driven Systems (EP/S021892/1). Additionally, the project stakeholders CDSM.

## 7 REFERENCES

[1] James Arbuckle and James H Nugent. 1973. A general procedure for parameter estimation for the law of comparative judgement. Brit. J. Math. Statist. Psych. 26, 2 (1973), 240–260.

[2] Tom Benton and T Gallagher. 2018. Is comparative judgement just a quick form of multiple marking. Research Matters: A Cambridge Assessment Publication (26) (2018), 22–28.

[3] Marie-Josée Bisson, Camilla Gilmore, Matthew Inglis, and Ian Jones. 2016. Measuring conceptual understanding using comparative judgement. International Journal of Research in Undergraduate Mathematics Education 2, 2 (2016), 141–164.

[4] T Bramley. 2007. Paired comparison methods. Techniques for monitoring the comparability of examination standards 246 (2007), 294.

[5] Tom Bramley. 2015. Investigating the reliability of adaptive comparative judgment. Cambridge Assessment, Cambridge 36 (2015).

[6] James AP Collingwood, Michael Wright, and Roger J Brooks. 2022. Evaluating the effectiveness of different player rating systems in predicting the results of professional snooker matches. European Journal of Operational Research 296, 3 (2022), 1025–1035.

[7] Tom Crick. 2021. COVID-19 and digital education: A catalyst for change? ITNOW 63, 1 (2021), 16–17.

[8] Tom Crick, Cathryn Knight, Richard Watermeyer, and Janet Goodall. 2020. The impact of COVID-19 and "Emergency Remote Teaching" on the UK computer science education community. In United Kingdom & Ireland Computing Education Research conference. 31–37.

[9] Justin Dillon and Meg Maguire. 2011. Becoming a teacher: Issues in secondary education. McGraw-Hill Education (UK).

[10] Arpad Elo. 1978. The rating of chess players, past and present (Arco, New York). (1978).

[11] R Michael Furr. 2021. Psychometrics: an introduction. SAGE publications.

[12] George A Gescheider. 2013. Psychophysics: the fundamentals. Psychology Press.

[13] Jason Hsu. 1996. Multiple comparisons: theory and methods. CRC Press.

[14] David R Hunter. 2004. MM algorithms for generalized Bradley-Terry models. The annals of statistics 32, 1 (2004), 384–406.

[15] MJ Kolen and RL Brennan. 2016. 'No More Marking': An online tool for comparative judgement. ISSN 1756-509X (2016), 12.

[16] Neil Marshall, Kirsten Shaw, Jodie Hunter, and Ian Jones. 2020. Assessment by comparative judgement: An application to secondary statistics and English in New Zealand. New Zealand Journal of Educational Studies 55, 1 (2020), 49–71

[17] National Education Union. 2021. workload and working time. Retrieved May 27, 2017 from https://neu.org.uk/advice/workload-and-working-time#:~:text=All%20teachers%20who%20teach%20pupils,time%20of%2020%20per%20cent.

[18] Radek Pelánek. 2016. Applications of the Elo rating system in adaptive educational systems. Computers & Education 98 (2016), 169–179.

[19] Alastair Pollitt. 2012. Comparative judgement for assessment. International Journal of Technology and Design Education 22, 2 (2012), 157–170.

[20] Research ED. 2018. Comparative judgement: the next big revolution in assessment? Retrieved May 27, 2017 from https://researched.org.uk/2018/07/06/comparative-judgement-the-next-big-revolution-in-assessment-2/

[21] Angela A Siegel, Mark Zarb, Bedour Alshaigy, Jeremiah Blanchard, Tom Crick, Richard Glassey, John R Hott, Celine Latulipe, Charles Riedesel, Mali Senapathi, et al. 2021. Teaching through a Global Pandemic: Educational Landscapes Before, During and After COVID-19. In Proceedings of the 2021 Working Group Reports on Innovation and Technology in Computer Science Education. 1–25.

[22] Connor Sullivan and Christopher Cronin. 2016. Improving elo rankings for sports experimenting on the english premier league. Virginia Tech CSx824/ECEx424 technical report, VA, USA (2016).

[23] Louis L Thurstone. 1927. A law of comparative judgment. Psychological review 34, 4 (1927), 273.

[24] Louis L Thurstone. 1927. Psychophysical analysis. The American journal of psychology 38, 3 (1927), 368–389.

[25] Kristi Tsukida and Maya R Gupta. 2011. How to analyze paired comparison data. Technical Report. WASHINGTON UNIV SEATTLE DEPT OF ELECTRICAL ENGINEERING.

[26] UK Government. 2021. Reducing school workload. Retrieved May 27, 2017 from https://www.gov.uk/government/collections/reducing-school-workload

[27] Richard Watermeyer, Tom Crick, Cathryn Knight, and Janet Goodall. 2021. COVID-19 and digital disruption in UK universities: Afflictions and affordances of emergency online migration. Higher Education 81, 3 (2021), 623–641.

[28] Richard Watermeyer, Kalpana Shankar, Tom Crick, Cathryn Knight, Fiona McGaughey, Joanna Hardman, Venkata Ratnadeep Suri, Roger Chung, and Dean Phelan. 2021. 'Pandemia': a reckoning of UK universities' corporate response to COVID-19 and its academic fallout. British Journal of Sociology of Education 42, 5-6 (2021), 651–666.

[29] Sally Weale. 2021. NEU Teacher Survey. Retrieved May 27, 2017 from https://www.theguardian.com/uk-news/2021/apr/08/one-in-three-uk-teachers-plan-to-quit-says-national-education-union-survey

[30] Jerry Wellington. 2007. Secondary education: The key concepts. Routledge.

[31] Christopher Wheadon, Patrick Barmby, Daisy Christodoulou, and Brian Henderson. 2020. A comparative judgement approach to the large-scale assessment of primary writing in England. Assessment in Education: Principles, Policy & Practice 27, 1 (2020), 46–64.